\documentclass[10pt,conference]{IEEEtran}
\IEEEoverridecommandlockouts
\usepackage{cite}
\usepackage{amsmath,amssymb,amsfonts}
\usepackage{algorithmic}
\usepackage{graphicx}
\usepackage{textcomp}
\usepackage{xcolor}
\newtheorem{definition}{Definition}
\def\BibTeX{{\rm B\kern-.05em{\sc i\kern-.025em b}\kern-.08em
    T\kern-.1667em\lower.7ex\hbox{E}\kern-.125emX}}
\begin{document}

\title{Modeling Language for Scenario Development of Autonomous Driving Systems
}

\author{\IEEEauthorblockN{Toshiaki Aoki}
\IEEEauthorblockA{
\textit{Graduate School of Advanced Science and Technology}\\
\textit{JAIST} \\
toshiaki@jaist.ac.jp}
\and
\IEEEauthorblockN{Takashi Tomita}
\IEEEauthorblockA{
\textit{Research Center for Advanced Computing Infrastructure}\\
\textit{JAIST} \\
tomita@jaist.ac.jp
}
\and
\IEEEauthorblockN{Tatsuji Kawai }
\IEEEauthorblockA{
\textit{Department of Computer Science}\\
\textit{Kochi University} \\
tatsuji.kawai@kochi-u.ac.jp}
\and
\IEEEauthorblockN{Daisuke Kawakami}
\IEEEauthorblockA{\textit{Mitsubishi Electric Corporation} \\
Kawakami.Daisuke@ab.MitsubishiElectric.co.jp}
\and
\IEEEauthorblockN{Nobuo Chida}
\IEEEauthorblockA{\textit{Mitsubishi Electric Corporation} \\
Chida.Nobuo@ab.MitsubishiElectric.co.jp}
}

\maketitle

\begin{abstract}
Autonomous driving systems are typically verified based on scenarios. To represent the positions and movements of cars in these scenarios, diagrams that utilize icons are typically employed. However, the interpretation of such diagrams is typically ambiguous, which can lead to misunderstandings among users, making them unsuitable for the development of high-reliability systems. To address this issue, this study introduces a notation called the car position diagram (CPD). The CPD allows for the concise representation of numerous scenarios and is particularly suitable for scenario analysis and design. In addition, we propose a method for converting CPD-based models into propositional logic formulas and enumerating all scenarios using a SAT solver. A tool for scenario enumeration is implemented, and experiments are conducted on both typical car behaviors and international standards. The results demonstrate that the CPD enables the concise description of numerous scenarios, thereby confirming the effectiveness of our scenario analysis method.
\end{abstract}


\section{Introduction}
Various scenarios surround automated vehicles; thus, it
is impossible to test all scenarios in advance. The National Highway Traffic
Safety Administration (NHTSA) has proposed a method to systematically
reduce the range of possible scenarios\cite{NHTSA}. In this method, the operational design domain (ODD) and object and event detection and response (OEDR) are first clarified, which represent the assumption of autonomous vehicles
and the tasks of automated vehicles, including 
object detection and appropriate responses, respectively. Test scenarios are then
created based on such ODD and OEDR. In the literature\cite{JAMA}, a
method to obtain scenarios based on variations in road structures and vehicle
motions has been proposed.

\begin{figure}
  \begin{center}
    \centerline{\includegraphics[scale=0.4]{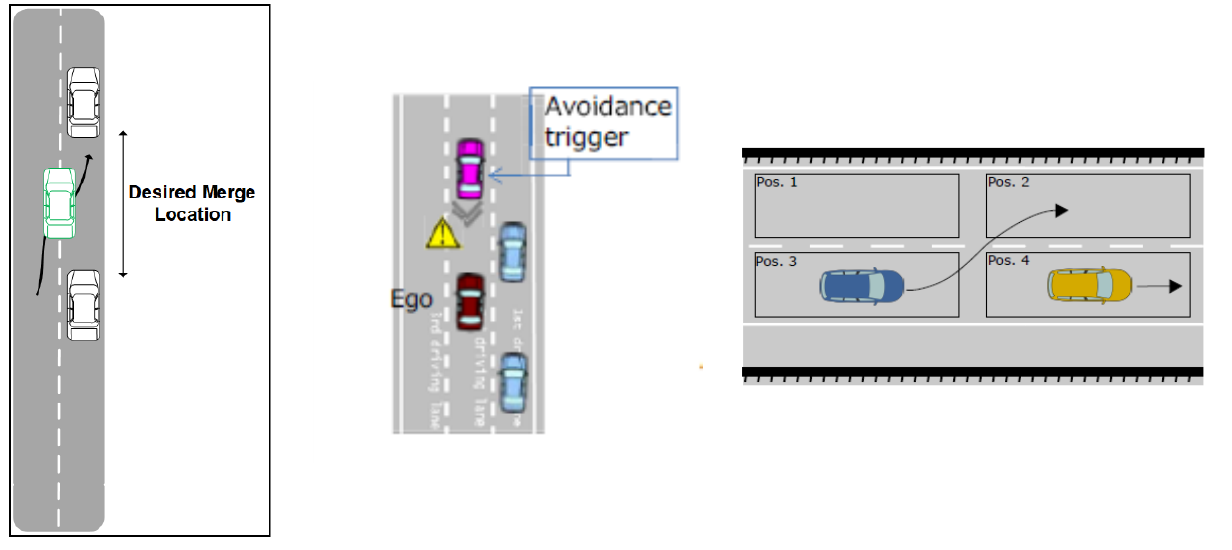}}
  \end{center}
  \label{ponchie}
  \caption{Scenario Description}
\end{figure}

The diagrams shown in Fig.\ref{ponchie} are typically used to
represent the location and behavior of vehicles in a scenario. These
diagrams, from left to right, were used in \cite{NHTSA},
\cite{JAMA}, and \cite{Bagschik}, respectively. Although such diagrams are
useful to intuitively understand scenarios, they are likely to cause
inconsistencies in viewers’ understanding due to their ambiguity. Therefore, they are not appropriate as notations for developing
highly reliable systems, such as automated driving systems.

In scenario analysis and design, it is crucial that notations not only
represent scenarios clearly and comprehensibly but also maintain the
complexity at a manageable level. Scenarios serve as the ultimate
under-approximation, making it essential to ensure that critical scenarios
are thoroughly covered. However, the number of scenarios can be very
large. As it increases because of the combinations of
elements, their interactions, and temporal sequences, 
comprehensive analysis becomes increasingly challenging. Therefore, 
notations must allow concise and efficient representation of 
numerous and complex scenarios without sacrificing clarity. Furthermore,
notations should facilitate the ease of interpretation, supporting both
human understanding and automated processes for tasks such as
verification, simulation, and scenario enumeration. To address these
issues, our approach is to compactly represent as many scenarios as possible. This allows for a visual review to confirm whether
critical scenarios are included.

In this study, we propose a graphical notation called the car position diagram (CPD) for scenario development. A single CPD model can represent numerous scenarios, and each scenario can be enumerated using a SAT solver. Furthermore, we propose a method for converting CPD models into propositional logic. By repeatedly solving the satisfiability of the converted formulas, we can comprehensively enumerate all scenarios. This transformation into propositional logic also defines the semantics of the CPD, providing translational semantics \cite{Meyer}. We conducted a series of experiments using the CPD to demonstrate the effectiveness of our approach. First, we modeled basic lane-changing patterns, which are fundamental behaviors of vehicles, and performed basic evaluations. Next, we artificially scaled the model to evaluate performance. In addition, we confirmed that the CPD supports practical scenario-based safety analysis using the automated driving safety evaluation framework proposed by the Japan Automobile Manufacturers Association (JAMA) and ISO 34502. The remainder of this paper is as follows: Section 2 reviews related research, Section 3 introduces the CPD, Section 4 discusses its formalization, Section 5 presents the experimental results, and Section 6 provides concluding remarks.

\section{Related Works}
In scenario-based methods, terms such as "scenario," "scene," and "situation" are frequently used, which makes it necessary to clarify their meanings. Ulbrich et al. \cite{Ulbrich} organized and defined the vocabulary related to scenarios, defining a scenario as a sequence of scenes arranged in a time-progress order, with each scene representing a snapshot of the environment, including elements such as background and dynamic objects. In this study, we adopt this terminology. Scenarios can exist at various levels of abstraction. Menzel et al. \cite{Menzel} categorized scenarios into three levels of abstraction depending on the development stage: functional, logical, and concrete. Functional scenarios describe road structures and dynamic objects using natural language, whereas logical scenarios specify states and parameters. Concrete scenarios define parameter values and generate executable test cases. Schuldt \cite{Schuldt} further decomposed scenarios into five layers: road-level (L1), traffic infrastructure (L2), temporary manipulation of L1 and L2 (L3), objects (L4), and environment (L5). An example from Bagschik et al. \cite{Bagschik} is shown on the right side of Fig. \ref{ponchie}. L1 represents two lanes and a hard sidewall, whereas L2 specifies solid white lines at the road boundaries and dashed white lines for lane markings. L3 is not represented. L4 illustrates a blue vehicle changing lanes from behind a yellow vehicle on the right lane to the left lane. L5, which is not shown in Fig. \ref{ponchie}, represents normal weather and temperature. The CPD focuses on modeling logical scenarios at the L4 level.

Unified formats for scenario data exchange in autonomous driving simulators and tools have also been proposed. Association for Standardization for Automation and Measuring Systems has introduced a format called OpenSCENARIO\cite{OpenSCENARIO}. Static elements such as road networks are represented using a format called OpenDrive\cite{OpenDrive}. Lanelets\cite{Lanelets} is a format that divides roads into polygonal sections, which allows road networks to be represented by combining these sections. There have also been attempts to extend map data such as Open Street Map\cite{OSM} using Lanelets. In addition, domain-specific languages for scenario descriptions have been proposed for similar purposes\cite{Zhang, Queiroz, Bock}.

Representative methods that visually represent vehicle position relationships include the traffic sequence chart \cite{TSC} and the method proposed by Back et al.\cite{Back}. The traffic sequence chart extends unified modeling language sequence diagrams to visually represent sequences of scenes, whereas Back et al.’s method lists continuous scenes, such as film frames. These methods allow for the visual representation of scenarios, thereby facilitating individual scenario analysis. However, they are inadequate for concisely representing numerous scenarios; thus, they are unsuitable for comprehensive scenario analysis.

Several methods have been proposed for generating scenarios for autonomous vehicles, including extracting scenarios from existing data \cite{Althoff}, using ontologies \cite{Bagschik}, and search-based testing \cite{Ben,Althoff2,Beglerovic,Khastgir,Hauer,Calo}. Althoff et al. \cite{Althoff} and Bagschik et al. \cite{Bagschik} generated scenarios based on past data and knowledge. Search-based testing methods define specific search spaces to extract relevant scenarios. Although these methods account for vehicle position information, they operate at a lower level of abstraction, which is closer to concrete scenarios. In contrast, our focus is on scenario modeling in the broader context of scenario development, analyzing scenarios at a higher level of abstraction. In addition, our approach is unique in that it generates scenarios based on models that are iteratively refined through human effort, including trial and error via manual review.

\section{Car Position Diagram}
\subsection{Overview}
\begin{figure}[h]
\begin{center}
\centerline{\includegraphics[scale=0.35]{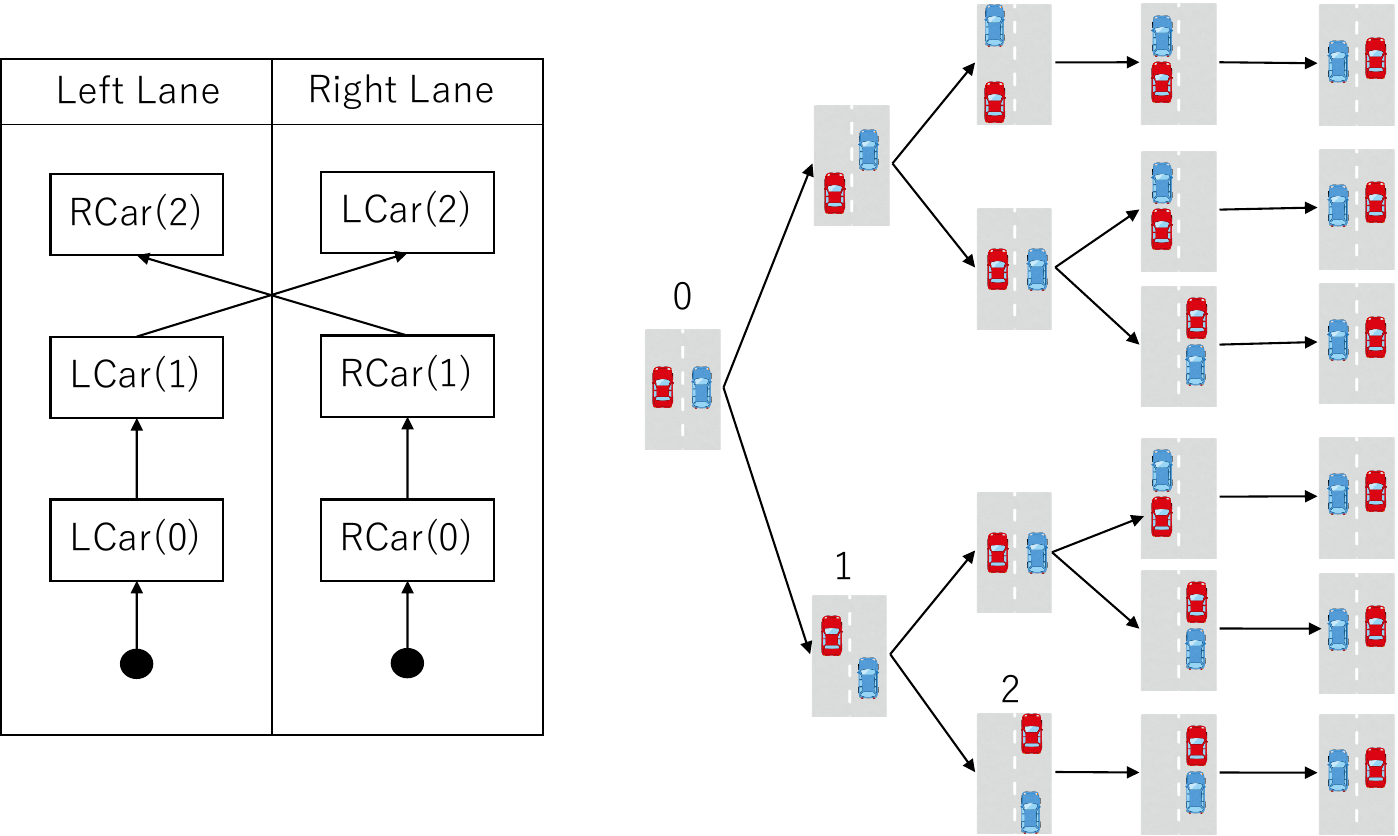}}
\end{center}
\caption{Example of CPD}
\label{example}
\end{figure}

An example of the CPD is shown on the left side of Fig. \ref{example}. The terms "Left Lane" and "Right Lane" represent the left and right lanes of the road, respectively. In this example, there are two cars: LCar and RCar. LCar(0), LCar(1), and LCar(2) represent the positions of LCar, which are referred to as boxes. Initially, LCar is at the position LCar(0). The arrows from LCar(0) to LCar(1) and from LCar(1) to LCar(2) indicate the movement of the car, which are referred to as box transitions. LCar moves from LCar(0) to LCar(1) and then to LCar(2). The same applies to RCar. On the right side of Fig. \ref{example}, the scenario represented by the CPD is shown as a tree structure. LCar and RCar are depicted as a red car and a blue car, respectively. Initially, as shown in 0, LCar and RCar run in parallel. Then, there are two possibilities: 1) where RCar moves ahead of LCar; 2) where LCar moves ahead of RCar.

This sequence of scenes, or scenario, can be obtained using the token-based approach of Petri nets. In the initial state, tokens are placed at LCar(0) and RCar(0). By extracting only the boxes containing tokens, Scene 0 can be formed. Next, when a token moves from LCar(0) to LCar(1), the tokens are placed at LCar(1) and RCar(0), forming Scene 1. If LCar’s token advances further, LCar moves to the Right Lane, forming Scene 2. By advancing the tokens step by step and forming scenes from the boxes containing the tokens, a scenario is constructed. In this example, as shown on the right side of Fig. \ref{example}, six scenarios are represented.

\subsection{Box Transitions}\label{BoxTransitions}
\begin{figure}
\begin{center}
\centerline{\includegraphics[scale=0.5]{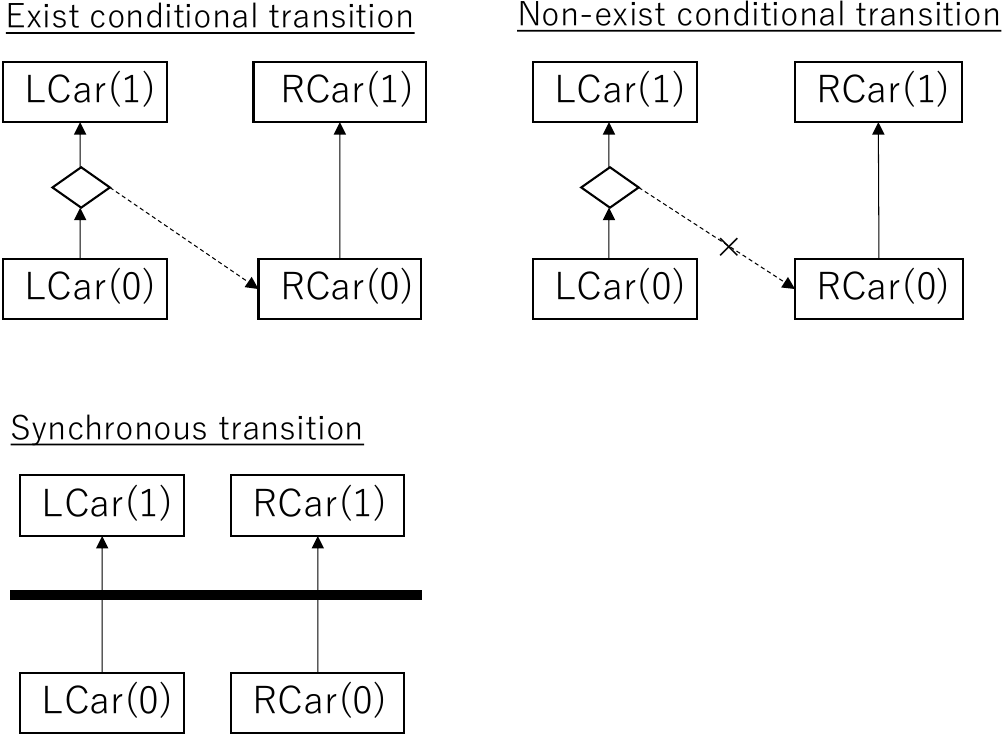}}
\end{center}
\caption{Transitions}
\label{transition}
\end{figure}

In addition to the standard transitions shown in Fig. \ref{example}, three other types of transitions exist (Fig. \ref{transition}): exist-conditional, non-exist-conditional, and synchronous transitions. In exist-conditional and non-exist-conditional transitions, other boxes are specified by dashed lines. In the former case, the transition fires when a token is present in the specified box, and in the latter case, the transition fires when no token is present. In the example shown on the top left of Fig. \ref{transition}, the transition from LCar(0) to LCar(1) fires when tokens are present in both LCar(0) and RCar(0). If there is no token in RCar(0), the transition will not fire. In the top middle example in Fig. \ref{transition}, the transition from LCar(0) to LCar(1) fires when there is a token in LCar(0) but no token in RCar(0). If there is a token in RCar(0), the transition will not fire. In synchronous transitions, multiple transitions fire simultaneously. In the bottom left example in Fig. \ref{transition}, when tokens are present in LCar(0) and RCar(0), two transitions fire simultaneously, and in the next step, the tokens are placed in LCar(1) and RCar(1). There are no scenes in which tokens are present in both LCar(0) and RCar(1) or LCar(1) and RCar(0).

\begin{figure}
\begin{center}
\centerline{\includegraphics[scale=0.5]{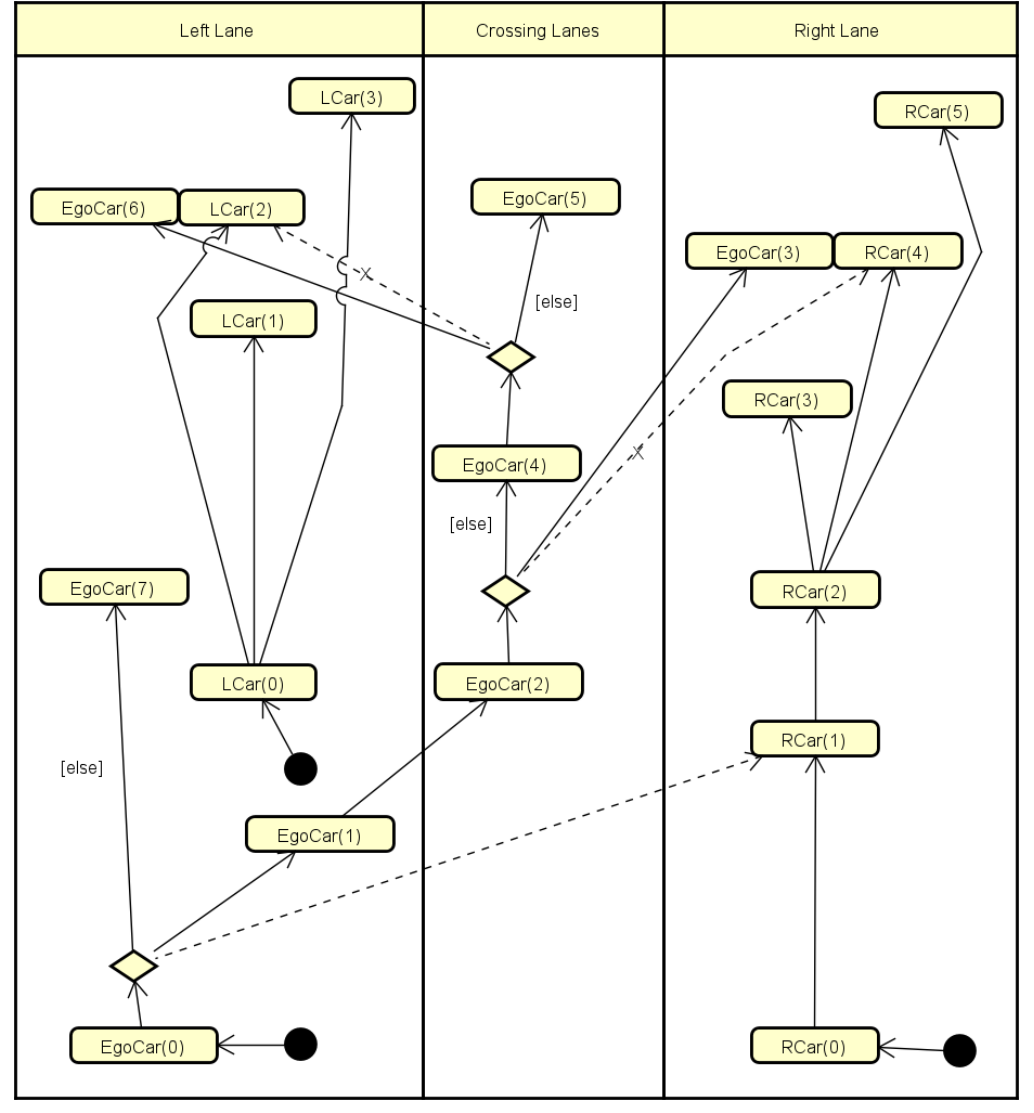}}
\end{center}
\caption{Lane change}
\label{lanechange}
\end{figure}

Fig. \ref{lanechange} shows an example of a lane change. "Left Lane," "Right Lane," and "Crossing Lane" represent the left lane, right lane, and a situation where the car is straddling both lanes. In this example, there are three cars, namely, LCar, RCar, and EgoCar, with EgoCar moving from the left lane to the right lane. The exist-conditional transition from EgoCar(0) to EgoCar(1) includes an alternative transition from EgoCar(0) to EgoCar(7), which fires when no token is present in RCar(1). This is the same as a non-exist-conditional transition but has been written in this way for clarity, a form of syntax sugar. EgoCar(6) and LCar(2), as well as EgoCar(3) and RCar(4), are at the same height in their respective lanes, indicating a potential collision.

\subsection{Metamodel}
\begin{figure*}[h]
\begin{center}
\centerline{\includegraphics[scale=0.55]{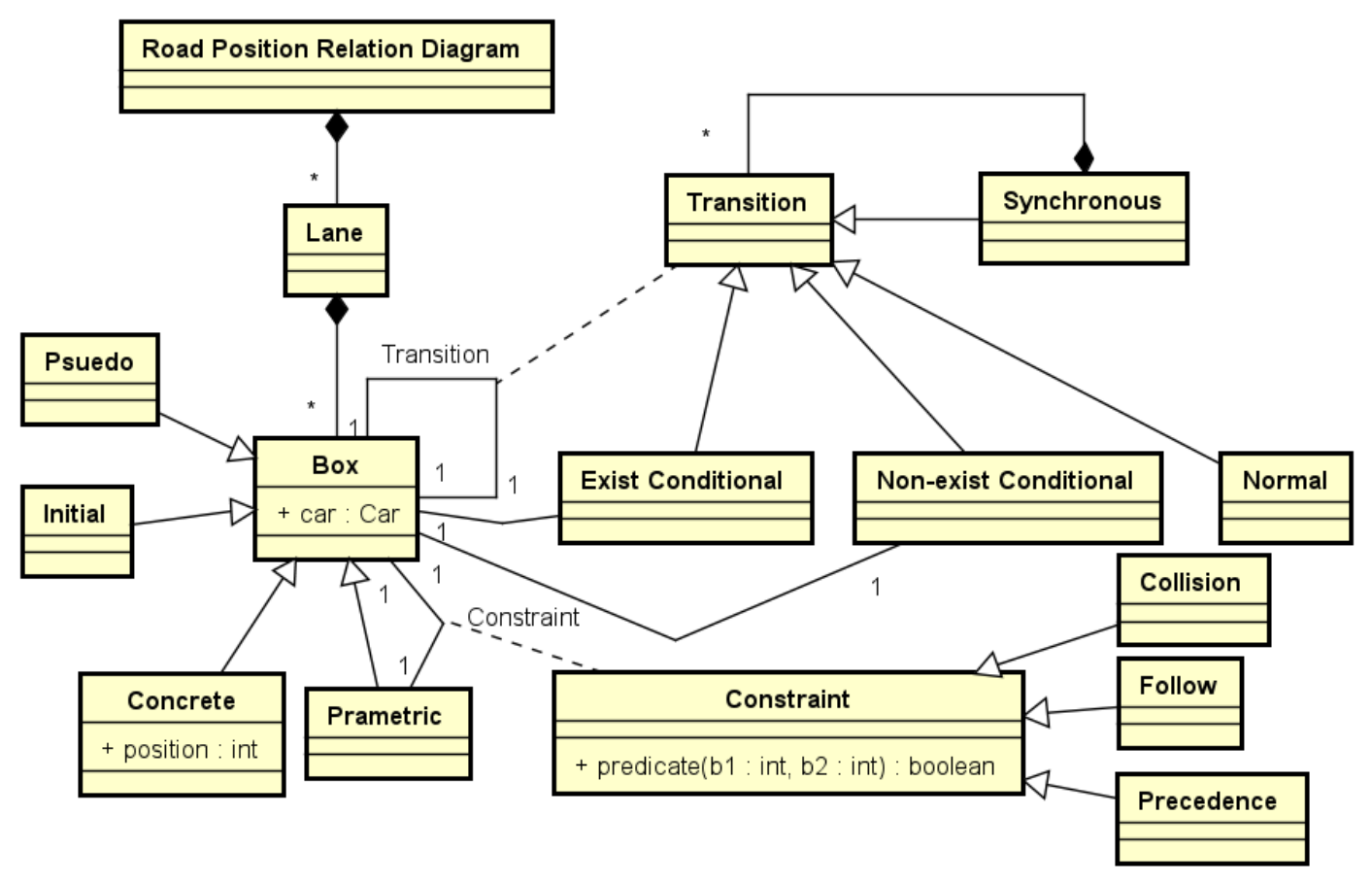}}
\end{center}
\caption{Metamodel}
\label{metamodel}
\end{figure*}

The metamodel of the CPD is shown in Fig. \ref{metamodel}. The CPD comprises multiple lanes, each containing boxes. There are two types of boxes: concrete and parametric boxes. In concrete boxes, an exact position is expressed as an integer, determined by the location of the box in the CPD. Let the position of each box $b$ be denoted as $Pos(b)$. For the boxes in Fig. \ref{example}, if they are concrete boxes, $Pos(LCar(0)) = Pos(RCar(0)) = 0$, $Pos(LCar(1)) = Pos(RCar(1)) = 1$, and $Pos(LCar(2)) = Pos(RCar(2)) = 2$. In contrast, parametric boxes express the position as a variable, and relationships with other boxes are given as constraints. These constraints include conditions such as collisions, following, and leading. If the boxes in Fig. \ref{example} are parametric boxes, the constraints include $Pos(LCar(0)) < Pos(LCar(1))$, $Pos(RCar(0)) < Pos(RCar(1))$, $Pos(LCar(1)) < Pos(LCar(2))$, and $Pos(RCar(1)) < Pos(RCar(2))$. Parametric boxes allow for flexibility in positioning. As discussed in Section \ref{BoxTransitions}, box transitions can be standard, exist-conditional, non-exist-conditional, or synchronous transitions, which are represented by the normal, exist-conditional, non-exist-conditional, and synchronous classes, respectively.

\section{Formalization of Car Position Diagram}
\subsection{Basic Elements}
In this section, we formalize the CPD. First, we define the basic elements of the CPD.

\begin{definition} Cars and Boxes

Let $Car$ be a set of identifiers representing cars. The set of rectangles representing the positions of cars is defined as $Box$, which is defined as follows:
 \[ Box = Car \times N\]
The position of a car $c \in Car$ at $(c, i) \in Box, i \in N$ is denoted as $c(i)$. In addition, the set of positions limited to car $c \in Car$ is denoted as $Box(c)$ and is defined as follows:
\[Box(c) = \{ (x,i) \in  Box | x = c \}\]
\end{definition}

\begin{definition} Transition Graph

The transition graph representing the movement of car $c$ is denoted as $G(c)$ and is defined as follows:
   \[
  \begin{array}{l}
    G(c) = (Box(c), E(c), b_0(c))\\
    \mbox{where}~ b_0(c) \in Box(c), \\
    E(c) = E_n(c) \uplus E_e(c) \uplus E_{ne}(c), \\
    E_n (c) \subseteq Box(c)\times Box(c), \\
    E_e (c)\subseteq Box(c)\times Box(c)\times Box, \\
    E_{ne} (c) \subseteq Box(c)\times Box(c)\times Box
  \end{array}
  \]
\end{definition}

$E_n(c), E_e(c), and E_{ne}(c)$ denote the sets of normal, exist-conditional, and non-exist-conditional transitions, respectively; $(b, b') \in E_n(c)$ denotes a box transition from $b$ to $b'$, $(b, b', b_c) \in E_e(c)$ denotes a box transition from $b$ to $b'$ that fires when box $b_c$ exists, and $(b, b', b_c) \in E_{ne}(c)$ denotes a box transition from $b$ to $b'$ that fires when box $b_c$ does not exist. $b_0(c)$ represents the initial box that a token exists initially.

\begin{definition} Structure of CPD

The transition graph representing the movement of $n \in N$ cars $c_1, \cdots, c_n$ is denoted as $G$ and is defined as follows:
  \[
  \begin{array}{l}
    G = (\{G(c_i)| 0\leq i \leq n \}, E_s)\\
    \mbox{where}~ E_s \subseteq 2^{Box\times Box} s.t. \forall E\in
    E_s ~ t\in E. \exists i.  t\in E_n(c_i)
  \end{array}
  \]
\end{definition}

$E_s$ denotes synchronous transitions; $E_s$ denotes a family of sets of box transitions, with its elements representing transitions that fire simultaneously.

\begin{definition} Car Position and Driving Lanes

Let $(X, \preceq)$ be a totally ordered set. The position of a car is represented by the function $Pos: Box \rightarrow X$. Hereafter, we assume the totally ordered set $X$ to be the natural numbers $N$, without loss of generality. The driving lane of a car is represented by the function $Lane: Car \rightarrow N$.

\end{definition}

\begin{definition} Scenario

The scenario for $n$ cars $c_i \in Car, 0 \leq i \leq n, n \in N$ is denoted as $\pi$. $\pi$ denotes a function of the following type:
  \[ \pi : N \rightarrow  \prod_{0\leq i \leq n} Box(c_i)\]
Each element $\pi(j), j \in N$ is called a scene.
\end{definition}

\subsection{Formalization of Box Transitions}
As shown in Definition 3, the CPD comprises a set of graphs. As described in Section III.A, each graph can be interpreted using tokens. Here, each graph contains a unique token; thus, each graph can be considered a subclass of Safe Petri Nets. Ogata et al. \cite{Ogata} proposed a method for encoding Safe Petri Nets into propositional logic formulas and using a SAT solver for reachability analysis. Based on their method, we convert the CPD into propositional logic formulas within a finite number of steps. Then, using a SAT solver, we perform reachability analysis and propose a method to enumerate all scenarios. The semantics of the CPD are defined by the set of scenarios it represents; thus, the proposed scenario enumeration method establishes the semantics of the CPD.

The scene $\pi(j) = (b_1, \cdots, b_n) = (c_1(m_1), \cdots, c_n(m_n)), j \in N$ for $n \in N$ cars $c_i \in Car, 0 \leq i \leq n$ is represented as a propositional vector $S = (s[b_1], \cdots, s[b_n])$. $s[b_i]$ is a proposition that is true if car $c_i$ is at position $b_i \in Box(c_i)$. The transition $t(c) \in E(c)$ from scene $S$ to $S'$ in the graph $G(c)$ for car $c$ is defined by the function $T_{t(c)}(S, S')$.

When $t(c)$ is a normal transition ($t(c) \in E_n(c)$),
\[
\begin{array}{l}
T_{t(c)}(S,S') =  \displaystyle\bigwedge_{b\in \bullet t(c)\backslash t(c) \bullet} s[b] \wedge
\bigwedge_{b\in \bullet t(c)} \neg s[b]' \wedge \\
\displaystyle \bigwedge_{b\in t(c)\bullet} s[b]' \wedge 
\bigwedge_{b\in Box(c)\backslash (\bullet t(c) \cup t(c)
  \bullet)} (s[b] \leftrightarrow s[b]')
\end{array}
\]
$\bullet t(c)$ and $t(c) \bullet$ denote the sets of pre-transition and post-transition states of the box transition $t(c)$. This notation follows the convention of Petri Nets. In the CPD, it is a set consisting of a single element for normal, exist-conditional, and non-exist-conditional transitions.

When $t(c) = (b, b', b_c)$ is an exist-conditional transition ($t(c) \in E_e(c)$),
\[
\begin{array}{l}
  T_{t(c)}(S,S') =  \displaystyle s[b_c] \wedge
  \bigwedge_{b\in \bullet t(c)} s[b] \wedge 
\bigwedge_{b\in \bullet t(c) \backslash t(c) \bullet} \neg s[b]' \wedge \\
\displaystyle \bigwedge_{b\in t(c)\bullet} s[b]' \wedge 
\bigwedge_{b\in Box(c)\backslash (\bullet t(c) \cup t(c) \bullet)}
(s[b] \leftrightarrow s[b]')
\end{array}
\]
 
When $t(c) = (b, b', b_c)$ is a non-exist-conditional transition ($t(c) \in E_{ne}(c)$),
\[
\begin{array}{l}
  T_{t(c)}(S,S') =  \displaystyle \neg s[b_c] \wedge
  \bigwedge_{b\in \bullet t(c)} s[b] \wedge 
\bigwedge_{b\in \bullet t(c) \backslash t(c) \bullet} \neg s[b]' \wedge \\
\displaystyle \bigwedge_{b\in t(c)\bullet} s[b]' \wedge 
 \bigwedge_{b\in Box(c)\backslash (\bullet t(c) \cup t(c) \bullet)}
(s[b] \leftrightarrow s[b]')
\end{array}
\]
 
When $t_1, \cdots, t_n, n \in N$ are synchronous transitions ($\{t_1, \cdots, t_n\} \in E_s$),
\[
\begin{array}{l}
  T_{t_1, \cdots , t_n} (S, S') =  \\
  \displaystyle\bigwedge_{b\in \bullet t_1} s[b] \wedge
  \bigwedge_{b\in \bullet t_1 \backslash t_1 \bullet} \neg s[b]'
  \wedge \bigwedge_{b\in t_1\bullet} s[b]' \\
  \wedge \cdots \wedge  \\
   \displaystyle\bigwedge_{b\in \bullet t_n} s[b] \wedge
  \bigwedge_{b\in \bullet t_n \backslash t_n \bullet} \neg s[b]' \wedge
  \bigwedge_{b\in t_n\bullet} s[b]' \wedge \\
   \displaystyle \bigwedge_{b\in Box\backslash
    (\cup_{0\leq i \leq n}(\bullet t_i \cup t_i \bullet))}
  (s[b] \leftrightarrow s[b]')
\end{array}
\]
 
The initial boxes for $n \in N$ cars $c_i \in Car, 0 \leq i \leq n$ are defined by the function $I(S)$.
\[
I(S) = \bigwedge_{0\leq i \leq n} s[b_0(c_i)] \wedge
\bigwedge_{b \neq b_0(c_i), b\in Box(c_i), 0\leq i \leq n} \neg s[b]
\]

The transition function for $n \in N$ cars is defined by the function $T(S, S')$.
\[
\begin{array}{ll }
T(S, S') = & \displaystyle \bigvee_{1\leq i \leq n, t(c_i)\in E(c_i)} T_{t(c_i)} (S, S')\vee \\
& \displaystyle \bigvee_{ts \in E_s} T_{ts} (S, S') \vee  D(S, S')
\end{array}
\]
 
 $D(S, S')$ represents the case where no transition is fired in the CPD and is defined as follows:
\[
\begin{array}{l}
  D(S, S') = \\
  \displaystyle\bigwedge_{b\in Box(c_i), 0\leq i\leq n} s[b] \leftrightarrow s[b]'\wedge \\
\displaystyle\bigwedge_{b \in \bullet t(c_i), t(c_i)\in E_n(c_i), 0\leq i\leq n}
\neg s[b] \wedge \\
\displaystyle \bigwedge_{(b,b',c)\in E_e(c_i), 0\leq i\leq n } \neg
(s[b]\wedge s[c]) \wedge \\
\displaystyle \bigwedge_{(b,b',c)\in E_{ne}(c_i), 0\leq i\leq n } \neg
(s[b]\wedge \neg s[c]) \wedge
\displaystyle \bigwedge_{t\in E_s}\bigvee_{b\in \bullet t} \neg s[b]
\end{array}
\]

The notation $\bullet t$ is used even though $t$ in the last line is a set of box transitions. $\bullet t$ represents the set of pre-transition boxes which contain those of each box transition.

Using the function $T(S, S')$, the $k \in N$ step transition is represented by the following propositional logic formula:
\[I(S_0)\wedge T(S_0, S_1) \wedge T(S_1, S_2) \wedge \cdots \wedge T(S_{k-1}, S_k) \]

By determining the satisfiability of this propositional formula, if it is satisfiable, its assignment represents a scenario. Let the satisfiable assignment of the proposition $s[b]$ be denoted as $\sigma[b]$. The constraint representing a different assignment is denoted as $\neg (\wedge_{b \in Box} s[b] \leftrightarrow \sigma[b])$. By repeating the satisfiability determination process and adding this constraint, all scenarios can be enumerated within a finite number of steps. Note that, according to Definition 2, the transition graph of a car may include loops. If loops are included, the number of scenarios becomes infinite, and it is not possible to comprehensively enumerate them in the form of scenarios, as described in Definition 5. To comprehensively enumerate scenarios, it is necessary to describe loops within scenarios or employ other methods. Conversely, if the transition graph does not include loops, all scenarios can be comprehensively enumerated in a finite number of steps.

\section{Experiments}
\begin{figure}
\begin{center}
\centerline{\includegraphics[scale=0.3]{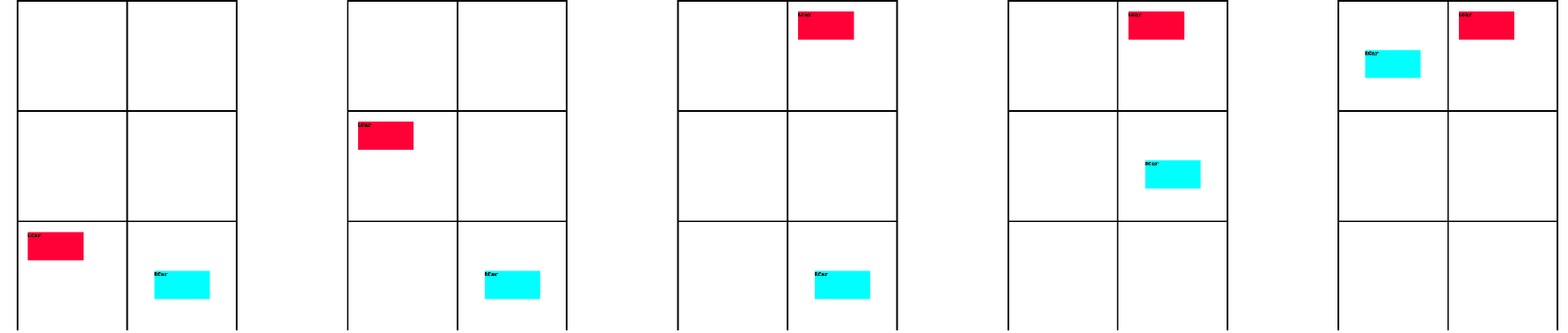}}
\end{center}
\caption{Generated Scenario}
\label{toolexample}
\end{figure}

We implemented a tool called GCPD that converts the CPD into propositional logic formulas and enumerates scenarios using Python and Z3py \cite{Z3}. This tool is available on GitHub \cite{CPD}. Figure \ref{toolexample} shows an example of a generated scenario, which
is one of the scenarios represented by the model in Figure \ref{example}.
Although the scenario is output as list data, it is visualized based on
that data.
Next, we describe experiments conducted using the tool.

\subsection{Lane Change Operation}
There are various autonomous driving behaviors, and a fundamental one is lane changing. For example, maneuvers such as overtaking, merging, and cutting in are combinations of lane changing. Therefore, we focused on lane changing and modeled different numbers of participating cars to enumerate the scenarios. The results are presented in Table \ref{result}. "Car" represents the number of cars, "Box" represents the number of boxes, and "Trans" represents the number of transitions (the numbers in parentheses, from left to right, are the counts of normal, exist-conditional, or non-exist-conditional transitions, and synchronous transitions). "Total" represents the total number of scenarios, and "Col" represents the number of scenarios where a collision occurs. The example in Fig. \ref{lanechange} corresponds to case 2-2 in Table \ref{result}. In this case, one Ego car and two principal other vehicles (totaling three cars) are participating. In case 2-2, 66 collision scenarios occur because RCar transitions from RCar(2) to RCar(4) after EgoCar transitions to EgoCar(3). By introducing synchronous transitions, collision-free scenarios can be created (case 2-1). Case 2-3 shows a situation in which all exist-conditional, non-exist-conditional, and synchronous transitions are converted to normal transitions. The symbols "nc," "c," and "n" following the case numbers represent cases where there are only collision-free scenarios, cases with (non) exist-conditional transitions but including collision scenarios, and cases where all transitions are normal, respectively. All CPD models are available in GitHub \cite{CPD}.

\begin{table}[h]
\caption{Results of lane changing modeling}
\label{result}
\begin{center}
\begin{tabular}{llllll}\hline
case &  Car &  Box & Trans & Total & Col \\ \hline
  1-1(nc) & 2 & 11 & 7 (3, 2, 2) & 4 & 0 \\
  1-2(n) & 2 & 11 & 9 (9, 0, 0) & 72 & 20 \\
  2-1(nc) & 3 & 18 & 13 (7, 4, 2) & 150 & 0 \\
  2-2(c) & 3 & 18 & 15 (9, 6, 0) & 522 & 66 \\
  2-3(n) & 3 & 18 & 15 (15, 0, 0) & 6480 & 1260 \\
  3-1(nc) & 4 & 22 & 15 (8, 4, 3) & 195 & 0 \\
  3-2(c) & 4 & 22 & 19 (9, 10, 0) & 1038 & 321 \\
  3-3(n) & 4 & 22 & 19 (19, 0, 0) & 169560 & 52240 \\ \hline
\end{tabular}
\end{center}
\end{table}

\begin{figure*}[t]
\begin{center}
\centerline{\includegraphics[scale=0.55]{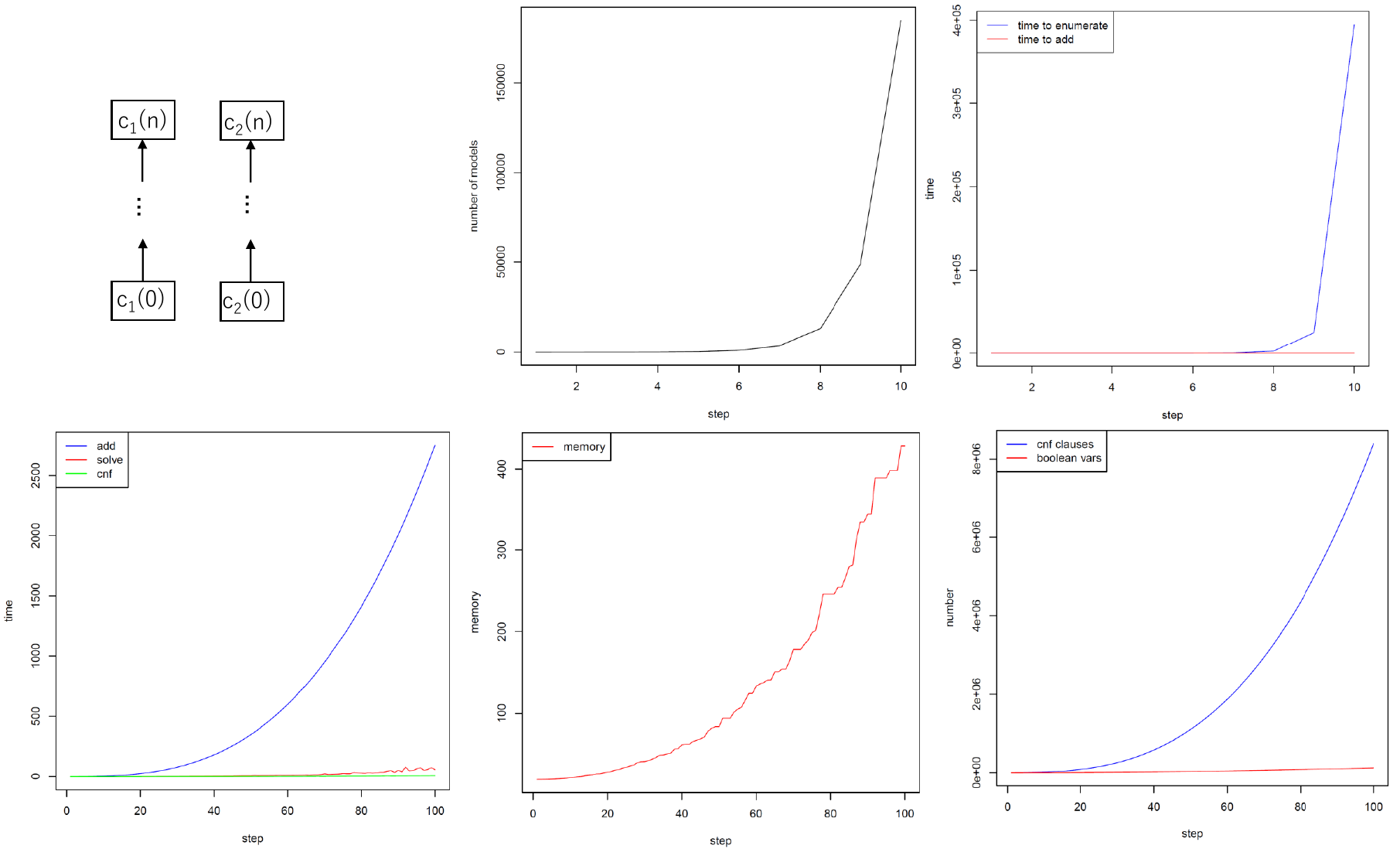}}
\end{center}
\caption{Performance}
\label{performance}
\end{figure*}

\subsection{Performance Evaluation}\label{PerformanceEvaluation}
We conducted experiments to evaluate the scenario generation performance. The PC used was an iMAC with a CPU of 3.6 GHz Intel Core i9 and 128 GB of memory. The model used is shown in the top left of Fig. \ref{performance}. Two cars are present, and each car has $n$ boxes. The number of generated scenarios was determined by increasing the value of $n$. The top left graph in Fig. \ref{performance} shows the relationship between $n$ and the number of generated scenarios. The top right graph shows the time required to enumerate the scenarios. The red line represents the time required to add constraints, and the blue line represents the total time required to enumerate all scenarios. The unit of time is s, which holds for the following figures. The horizontal axis represents the value of $n$. The bottom row of Fig. \ref{performance} shows the results for generating a single scenario. The left graph shows the time required to generate a scenario (in seconds). The blue line represents the time required to add constraints, the red line represents the time for satisfiability determination, and the green line indicates the time for conversion to the conjunctive normal form (CNF). The middle graph shows the memory usage (in MB), and the right graph shows the number of propositions (red line) and the number of CNF clauses (blue line). The conversion to the CNF is performed via Tseitin transformation \cite{Tseitin}.

As shown in Fig. \ref{performance}, the computation time for scenario generation increases significantly as the model size increases. For the model shown in the top left panel of Fig. \ref{performance}, the number of scenarios can be theoretically determined as $\frac{(2n)!}{(n!)^2}$. The number of scenarios increases factorially with respect to $n$, and the computation time increases accordingly. For $n=10$, the number of scenarios is 184,756, and the time required to enumerate them is 394,890.606019 s (approximately 4.57 days). Generating a single scenario does not require a significant time. For $n=100$, the computation requires 2807.6733 s. Most of this time is spent adding constraints. In the prototype tool, constraints are generated using the method described in Section 4, which is implemented in Python. As $n$ increases, the number of times the function $T(S,S')$ requires to be expanded increases, resulting in larger constraint formulas. This leads to a longer time for constraint addition in Python. Memory usage is not a significant issue, because the amount of memory consumed is relatively low. However, compared to the number of propositional variables, the number of CNF clauses is very large. For $n=100$, there are 120,809 propositional variables and 8,403,006 CNF clauses. Nevertheless, the time required for CNF conversion and satisfiability determination is significantly shorter than the time required for constraint addition. Therefore, the size of constraints is the primary factor affecting the time required to generate a single scenario, and the number of scenarios primarily influences the time required for scenario enumeration.

\begin{figure}
\begin{center}
\includegraphics[scale=0.4]{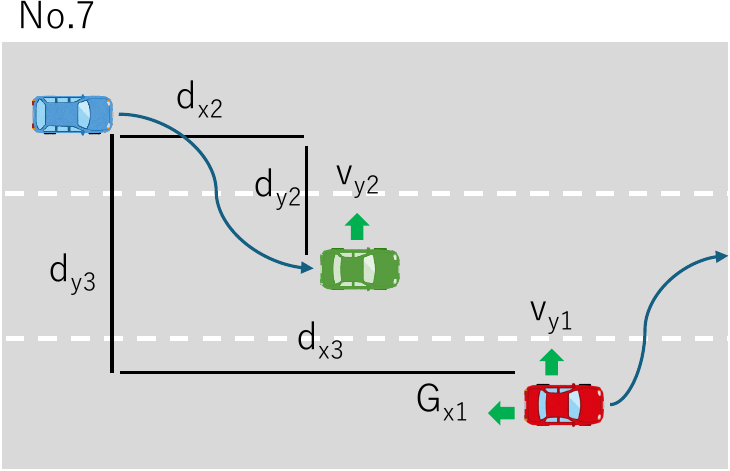}
\includegraphics[scale=0.5]{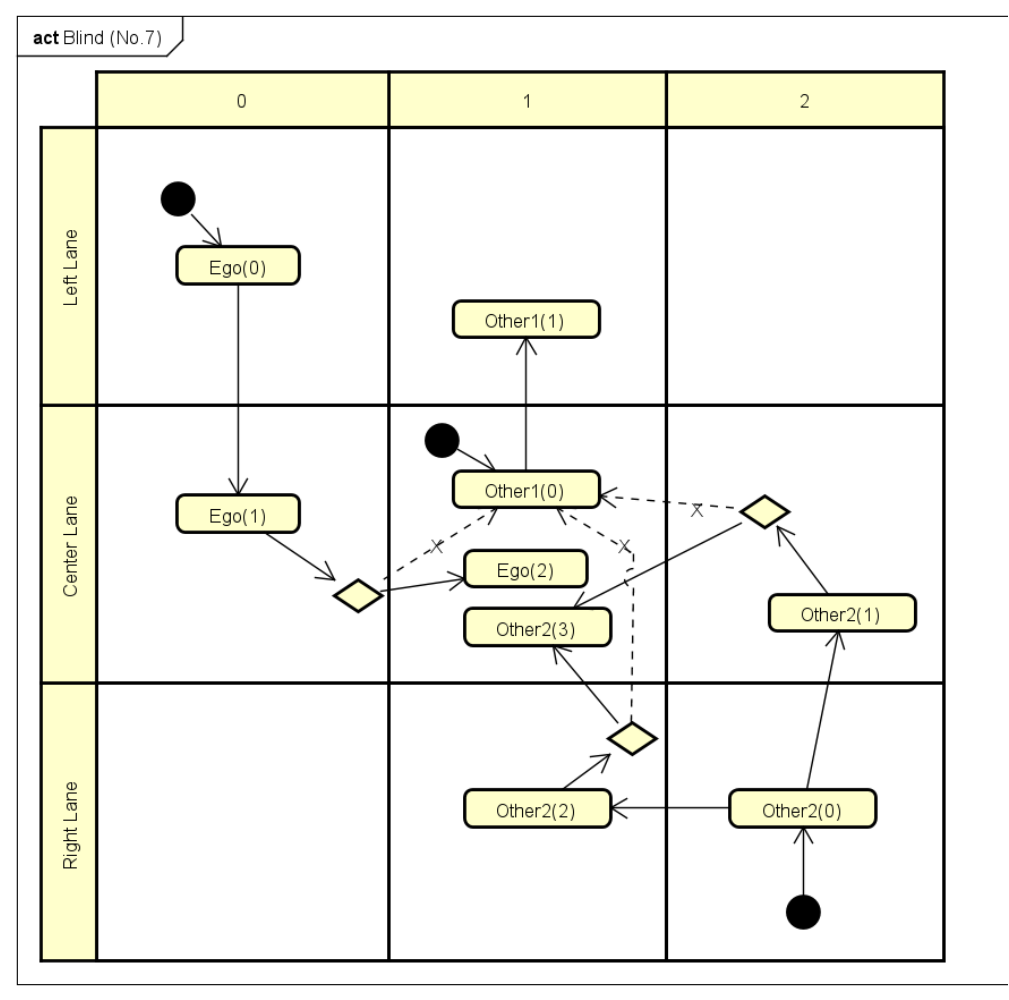}
\end{center}
\caption{Scenario No.7 - JAMA Framework}
\label{JAMAFramework1}
\end{figure}

\subsection{Case Studies}
\subsubsection{JAMA Framework}
JAMA has proposed the
automated driving safety evaluation framework, a scenario-based method for
safety analysis, hereafter referred to as the JAMA Framework. This framework
offers a structured approach for assessing the safety of autonomous driving
systems by evaluating various cognitive disturbances and interactions
between vehicles.  The JAMA framework is designed to facilitate 
rigorous evaluation of autonomous vehicles, ensuring that they meet high safety
standards before deployment. The framework is built on best practices and
systematically examines cognitive, traffic, and motion disturbances that
 automated driving systems may encounter. The framework first organizes
variations in road geometries, which involves the organization of different
types of road structures and layouts to simulate diverse driving
environments. Second, variations in surrounding car positions are considered to examine how different placements and movements of nearby
cars impact autonomous cars.  In addition, whether the surrounding
cars affect the ego car is assessed, focusing on interactions
between the ego car and other traffic participants. Finally, these
factors are integrated into comprehensive matrices, which form the basis for
scenario analysis.  In this case study, we demonstrate the application of
the CPD to model and analyze scenarios derived from the JAMA framework.

One example is Scenario No. 7, depicted at the top of Fig. \ref{JAMAFramework1}, which examines cognitive disturbances caused by blind spots. In this scenario, three cars are involved: blue, green, and red cars. The red car is positioned in the blind spot of the blue car. The sequence of events begins with the blue car moving from the left lane to the center lane, followed by the green car’s movement from the center lane to the left lane. Simultaneously, the red car moves from the right lane to the center lane. Using the CPD, we formalized the relative positions of the ego car and the other cars in the environment, as shown at the bottom of Fig. \ref{JAMAFramework1}. The blue, green, and red cars are represented as Ego, Other1, and Other2,
respectively. From the viewpoint of Ego, Other1 is visible, but Other2 is
not. Therefore, in the transition from Ego(1) to Ego(2), the non-existence of
Other1 is checked, but that of Other2 is not. Similarly, from
the viewpoint of Other2, Other1 is visible, but Ego is not. Thus, in the
transition from Other2(1) and Other2(2) to Other2(3), the non-existence of
Other1 is checked, but that of Ego is not.
The CPD formalization enabled us to identify 32 potential collision scenarios, which were categorized as follows:
\begin{itemize}
 \item 24 collisions caused by the ego car’s actions.
\item 16 collisions triggered by changes in acceleration.
\item 8 collisions due to deceleration of surrounding cars.
\item 8 collisions resulting from another car’s actions.
\end{itemize}

\begin{figure*}
\begin{center}
\centerline{\includegraphics[scale=0.4]{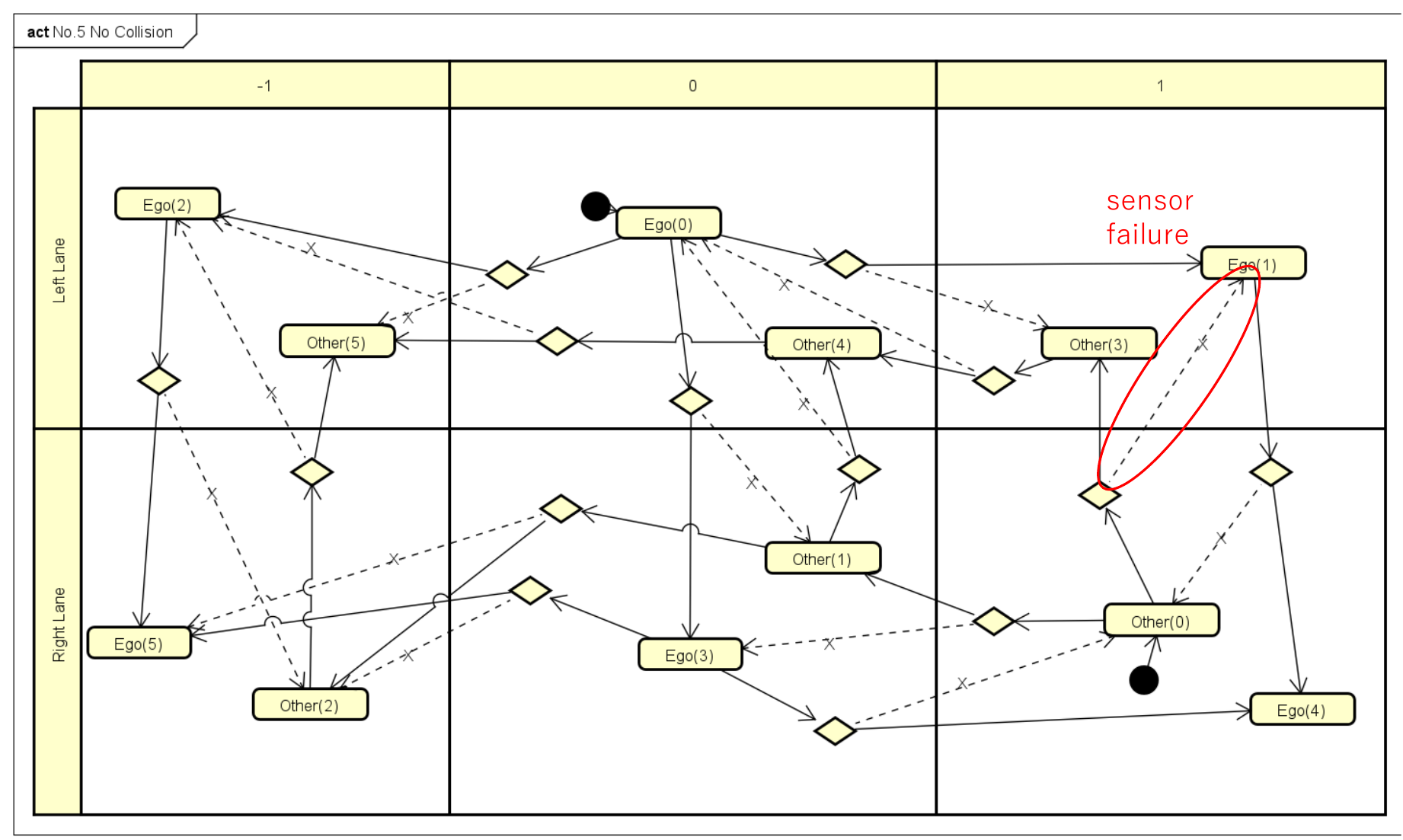}}
\end{center}
\caption{ Scenario 5JAMA Framework}
\label{JAMAFramework2}
\end{figure*}

Fig. \ref{JAMAFramework2} illustrates another scenario from the JAMA framework, focusing on cognitive disturbances, specifically sensor failures, as described in Scenario No. 5. In this scenario, two cars are involved, each occupying one of two lanes: the ego car on the left lane and another car on the right lane. Both cars change lanes simultaneously. Under normal conditions, where no sensor failure occurs, 47 no collision scenarios are identified. Since the sensor checks that no other car exists at the destination, it is represented by a non-exist-conditional transition. A sensor failure is modeled by replacing the non-exist-conditional transition with a normal transition. If a sensor failure occurs during lane changing (indicated by the red circle), three collision scenarios emerge.

By leveraging the automated reasoning capabilities of the CPD, we thoroughly enumerate and analyze these scenarios, ensuring that all potential collision cases are considered. The results demonstrated that visual ambiguities inherent in the JAMA matrices are resolved, allowing for more precise analysis of safety-critical interactions.

\subsubsection{ISO 34502}
ISO 34502 uses graphical notations named zone graphs.  A zone graph is a
conceptual model designed to simplify the analysis of complex traffic
situations for highly automated driving systems. A zone graph abstracts road layouts
and the interactions between cars and other road users, providing a
a clear structure that captures the dynamics of different traffic
scenarios. Instead of focusing on the exact road geometry, a zone
graph divides the environment into zones, each representing a distinct
part of the road or interaction.  These zones are interconnected by paths that
describe a car’s intended movements, potential conflicts with other
traffic participants, and the flow of information required for decision making. Variations in which traffic participants exist in zones are
represented by Zwicky-Boxes. A Zwicky-Box is a method used to explore and
organize all possible outcomes in a complex situation by decomposing them 
into key factors or variables. It works by considering each relevant aspect of
a situation, such as the conditions or decisions, and identifying
different options for the aspect. These options are then combined in
various ways to create a comprehensive map of all potential scenarios.

\begin{figure*}
\begin{center}
\centerline{\includegraphics[scale=0.7]{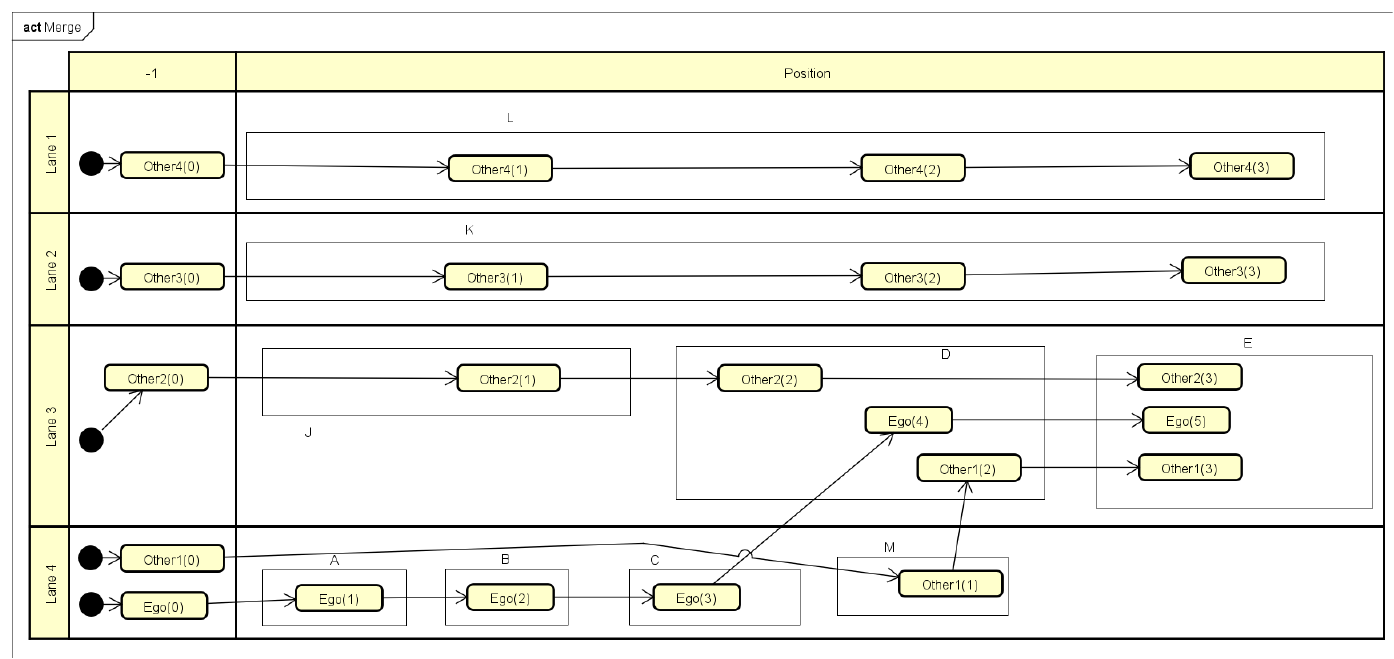}}
\end{center}
\caption{ISO 34502 Annex E}
\label{ISO34502}
\end{figure*}

In this case study, the CPD is applied to formalize and analyze the scenarios described in Annex E of ISO 34502, with a focus on a car merge situation. In this scenario, the ego car must decide whether to merge into a lane occupied by other cars. The scenario involves five cars, including the ego car, and four lanes. The analysis provided in Annex E offers guidance on defining zone graphs for such scenarios. Fig. \ref{ISO34502} illustrates the scenarios modeled using the CPD-based on zone graph and Zwicky-Box representations. The original zone graph contains nine zones labeled A, B, C, D, E, J, K, L, and M, and the corresponding boxes are depicted in the figure.

Due to the extensive number of scenarios in this case, the total count could not be determined within 1 h of computation; however, the number of identified collision scenarios was 3,600. Annex E defines two equivalence classes, “passable” and “not passable,” to indicate whether the ego car can safely merge into the lane. These equivalence classes group the positions of cars accordingly. There are three positions, Ego(1), Ego(2), and Ego(3), where the passable or not passable status is assigned. Each variation was defined as a constraint, and the corresponding scenarios were extracted immediately. The CPD enabled efficient identification of key scenarios within specific equivalence classes. However, the total number of scenario variations could not be fully enumerated.

The CPD shown in Fig. \ref{ISO34502} includes numerous scenarios; thus, it is impractical to list all. It is also unrealistic to perform testing or analysis on every scenario. Therefore, the task is to identify critical scenarios from the large set. In this case study, we extracted scenarios by combining equivalence classes and the positions of the ego car. Therefore, by defining constraints that serve as queries over numerous scenarios, representative scenarios can be obtained efficiently.

\section{Discussion}
As shown in Table \ref{result}, the CPD can concisely represent numerous scenarios. In cases 3-1, 3-2, and 3-3, where four cars are present, between 195 and 169, 560 scenarios can be represented using 22 boxes and 15–19 box transitions. The CPD allows for the compact representation of many scenarios, thereby enabling visual review. The number of scenarios represented by the CPD is vast, and for tasks such as analyzing the impact of adding (or removing) conditional transitions, a mechanism for enumerating scenarios is indispensable. In this study, we have also proposed a method for automatically obtaining comprehensive scenarios using a SAT solver.

When the number of scenarios is large, simply enumerating them makes it difficult to verify the presence of unintended scenarios. In case 2-2, non-exist-conditional transitions ensure that EgoCar does not change to the right lane when RCar is present at a collision point. The same applies to left-lane changes. However, in this case, 66 scenarios include collisions. Such collisions can be avoided by using synchronous transitions to make EgoCar and RCar transition simultaneously, which corresponds to case 2-1. In case 2-1, 150 scenarios are represented, which is too many to verify manually. In cases 3-1 and 3-2, the number of scenarios is even higher. Because the CPD uses a SAT solver for enumeration, it is sufficiently flexible to add as many constraints as required. Therefore, constraints expressing the expected and unexpected properties for scenarios can be added as appropriate. This allows model verification and scenario filtering. Proposing a language to describe such properties is a topic for future work.

As discussed in Section \ref{PerformanceEvaluation}, performance is primarily influenced by the number of scenarios. The number of scenarios generally increases factorially with the number of box transitions; however, this can be mitigated by adding constraints. As shown in Table \ref{result}, for the same number of cars, case (nc) has the fewest scenarios, followed by case (c), and case (n) has the most. Thus, adding constraints such as synchronous or non-exist-conditional transitions can reduce the number of scenarios. Additional constraints related to relative distances can also be considered. For example, in the example depicted in Fig. \ref{lanechange}, the combination of EgoCar(4) and RCar(5) is important, whereas the combination of EgoCar(0) and RCar(5) is less important. Therefore, it is possible to enumerate only scenes where the distance between boxes is within a certain range. In the example depicted in Fig. \ref{performance}, if the distance between each adjacent box is set to 1 and the scenes are limited to those where the box distance is less than 3, the number of scenarios for $n=10$ becomes 39,366. Without this limitation, the number of scenarios is 184,756, resulting in a reduction of approximately 79\%. Therefore, the number of scenarios can be controlled by adding various constraints.

The application of the CPD to both the JAMA and ISO frameworks presents significant advantages, particularly in terms of improving scenario formalization and ensuring reliability, which is critical in the context of safety. Within the JAMA framework, each matrix element represents a scenario; however, ambiguities exist due to the visual notations used in the documentation. In contrast, ISO 34502, which is derived from the JAMA framework, uses a more formal visual representation known as a zone graph. Although zone graphs improve upon the notations used in JAMA, they can be further refined, particularly regarding the incorporation of properties such as those found in Zwicky-Boxes and combinations of car locations.
Using the CPD, we successfully formalized scenarios from these standards, achieving a higher degree of precision in their analysis. The CPD enables more rigorous analysis, allowing specific properties to be validated against scenarios and supporting scenario enumeration. For example, within the JAMA framework, we systematically enumerated scenarios based on variations in collision cases. This approach is not limited to the JAMA framework and can be extended to other standards such as the New Car Assessment Programme)\cite{NCAP}, which focuses on collisions.
However, the merge scenario in ISO 34502 encompasses an extremely large number of possible scenarios; thus, full enumeration is infeasible under reasonable time constraints. Nevertheless, the CPD enabled efficient identification of key scenarios within specific equivalence classes. Using GCPD, scenarios can be mined by imposing constraints, because it employs a SMT solver to facilitate enumeration. To further enhance this capability, we plan to develop a query language that can extract critical scenarios from CPD models with greater efficiency.
Another potential application of the CPD lies in its use as a foundation for scenario coverage. As noted previously, the CPD provides a compact representation of numerous scenarios, thereby offering the possibility of establishing coverage criteria based on the CPD model. An initial approach involves covering all elements of the model, such as individual boxes, transitions, and combinations of n-boxes. However, further research is required to identify the coverage criteria that are most effective in practical implementations.

\section{Conclusion}
In this study, we have proposed the CPD, a graphical notation for scenario development aimed at verifying autonomous driving systems. The CPD is specifically designed for analyzing and designing logical scenarios at the L4 level. Although these scenarios are highly abstract, their number tends to be immense. Scenarios serve as the ultimate under-approximation, decomposing into specific cases; thus, it is crucial to ensure that critical scenarios are thoroughly covered. To achieve this, it is important to compare multiple scenario options and model them iteratively through trial and error. However, the large number of scenarios can make it challenging to effectively review them. The CPD addresses this issue by providing a compact visual representation of various scenarios. Furthermore, by utilizing a SAT solver, we can enumerate all scenarios represented by the CPD model.

As demonstrated in the experiments using the JAMA framework, we 
concisely analyzed scenario variations, such as collision
directions. However, as demonstrated in the experiments with ISO 34502, the total
number of scenarios remains very large. Although enumerating all scenarios is
challenging, we rapidly identified representative
scenarios. This indicates that by properly categorizing scenarios, e.g.,
using equivalence classes, representative scenarios can be
automatically extracted. In other words, we effectively mined important
scenarios. Given the ability of the CPD to represent numerous
scenarios on a manageable scale, it can serve as a foundation for
scenario mining.

We have focused on highly abstract scenarios. In future work, we plan to explore methods for integrating CPD models with concrete data,
such as map data \cite{OSM, Lanelets} and traffic data \cite{NGSIM}. This
will enable scenario modeling and analysis based on actual road and
traffic conditions. Furthermore, we plan to support more complex road
network topologies in future developments.

\section*{Acknowledgement}
This work was supported by JST, CREST Grant Number JPMJCR23M1, Japan.

\end{document}